\documentclass[showpacs,floatfix,fullpage,12pt]{article}

\usepackage{graphics}
\usepackage{graphicx}
\usepackage{subeqn}
\begin{document}
\begin{flushright}
TIFR/TH/06-24 \\
December 2006
\end{flushright}

\medskip

\pretolerance=10000

\begin{center}
\large{Coexisting Order in the Pseudogap State of Cuprates} \\ [5mm]
%\author{{\bf P. Bhattacharyya}} \\ [7mm]
{{\bf P. Bhattacharyya}} \\ [7mm]
%\affiliation{Department of Theoretical Physics, Tata Institute of 
%Fundamental Research,\,  Colaba, Mumbai 400005, India}
{Department of Theoretical Physics \\ 
Tata Institute of Fundamental Research, \\  
Colaba, Mumbai 400005, India}
\end{center}

\bigskip

\begin{abstract}
\vskip 0.2cm 
A pseudogap is shown to be a magnetic diffuson (MD) in a state with
classical localization order coexisting with Quantum Peierls (QP)
order. A soft quantum localization mode, a phason, ensures scale
invariance with strong correlations among different momentum states
seen as a circular Fermi surface with a gap in ARPES when phase
fluctuations are random. These correlations imply a state with
coexisting order parameters at distinct length scales. Charge
quantization and scale invariance accounts for the sharp peak in the
excess tunneling current at $V=0$.
\end{abstract}

\noindent PACS numbers: 74.90.+n, 75.20.-g, 74.20.Mn, 73.43.Cd

\bigskip

%\maketitle

%====================
%\section{Introduction}
%====================

An electronic mechanism of superconductivity [1-18] requires pairing
interactions mediated by spin or charge density fluctuations in the
presence of a gap in the excitation spectrum due to short range order
(SRO) [5,11,12].  An unusual feature of 2+1
dim. cuprates is $Cu-O$ bond adjustments which enable a resonant
tunneling of coherent local energy in a Quantum Peierls (QP) order SRO
state with the concomitant excitation of a magnetic diffuson
(MD). This, we believe, is the Higgs of the pseudogap state with a
relaxation rate which is linear in temperature or gap [6] since it
involves a small region of the Fermi surface (FS) with a weak phase
decohering length as seen in Infrared conductance [6]. This is
different from tunneling in conventional superconductors where the
tunneling is proportional to the square of the gap amplitude. 
Several important and experimentally
relevant theories of the pseudogap state are available. In the
d-density wave (DDW) theories [4], order parameter should have Ising
symmetry since they seek to describe SRO [5] which would become
unstable with low energy electronic excitations [6]. Neutron
diffraction experiments [7] have confirmed hidden order with
translational invariance as have thermal expansion [8] implying a soft
mode. Similarly RVB [9] and theories involving simultaneously solving
for a bound state t-matrix and superconducting order [11] should not
see a pseudogap resonsible for magentic relaxation [10] since only a
real gap of SRO will be formed [5,11,12]. Once a charge gap is
established, magnetic relaxation [10] and similar phenomena remain
open problems. Varma's phase fluctuation theory [2] makes it difficult
to explain a stable order parameter. Similarly, Kohn singularities
weakened by disorder at the superconducting insulating transition have
been proposed [20,21]. However, then, pseudogap state should be
antiferromagnetic (AF) not DW with a spectral function peaked at
$\omega = Q_c$ (MD wave vector) not $\omega = 0$ since spins are on
$Cu$ sites where anomalous NMR is observed.

Our proposal is a coherent transfer of a single electron with
excitation of a magnetic diffuson (MD) in the singlet state results in
a Kondo spin singlet (dyon) with a resonant level at a dipolar energy
with a coherent phase transfer of $\pm \pi/m (m=2$ s-wave, $m=3$
p-wave, $m=4$ d-wave order). Weak localization order at $k_{th} (T)$
with a quantized phason at $k_{th}$ results in Fermi arc lengths
$\propto (T/T_g)$ [14]. This in our opinion, is the cause of free
electrons at the Fermi surface [14,15,16] as in the strong coupling
regime of a Kondo spin. When superconductivity sets in, pairs of
electrons are coherently exchanged making the pseudogap a real gap
with phase coherence over the sample. Scale invariance requires SRO
parameter for a Quantum Peierls (QP) state at $Q \propto \pm (1/2)((1,
0)\ {\rm or}\ (0, 1))$ modulated by a oxygen-concentration order
parameter $k_x (\alpha x)$ giving SRO at $Q_\pm = Q \pm k_x$ in the
pseudogap state. Non-stoichiometry results in a voltage field along
the c-axis due to mode softening and a consequent modulation at $Q_v
\propto k_x$. For d-wave order symmetry, electron propagation is
allowed when $Q_v = k_x$ for gapless electrons at the Fermi surface, a
decreasing to zero at low temperatures in the superconducting
state. Complete order parameter $Q^\ast$ of the pseudogap state is
valued in $G = SU(4)$ as a direct sum of order parameters (Casimirs)
for correlations at different length scale as
\begin{equation}
Q^\ast_\alpha = Q_{p\alpha} \oplus Q_c \oplus  k_{v} \oplus k_{th} 
\end{equation}
where $Q_{p\alpha} (\alpha = \pm)$ is a Quantum Peierls (QP) order
parameter, $Q_c$ is a magnetic dyon or pseudogap and $Q_v$ (or $k_v$) is a weak
dipolar quantum order which is a voltage modulation causing a Raman
mode [2]. Finally, $k_{th} \alpha T$ is a (weak localization
length)$^{-1}$ responsible for the dephasing of circulating currents
and several interesting anomalies including a zero bias anomaly in I-V
characteristics [22] and a correlated phason gap with d-wave
symmetry seen as gaps states on the Fermi surface [14]. If
electromagnetic energy is, in fact, correlated with (de-) twinning
sites which are distributed randomly in the lattice, their phasons
will be scaled in units of a soft mode of weak localization resulting
in Stripes ordering at short distances as in some CDW's with a
modulating diffusonic order $Q_c$ in units of the soft mode energy. 
This is a manifestation of a hidden gauge
invariance [3] seen in spin glasses coming from permutations of localized
regions expected in a disordered state with a conserved current
\begin{equation}
J_{va} = Q \oplus a Q_v, \ \ a = \pm 
\end{equation}
with $J_{v+} J_{v-} = 1$ as required for SU(2) or SL(2,{\bf R})
invariance representing algebra closure with $\oplus$ representing a
factor of 1,$i$ or $\tau$ where $\tau^2 = 1 + \tau$. Then $Q_p \propto
(Q_v, Q_h)$ or $(Q_p, Q_h) \cdot Q_v = Q^2_c$ from modular $T$ duality
implying $Q_c$ is Casimir of EM duality being a magnetic dyon in a
state with a spontaneously broken orbital symmetry with a Higgs which
is the charge gap. $Q_v$ is a weak localization gap while $Q_c$ is the
magnetic diffuson (MD) Higgs mode restoring conformal scaling.
A random exchange of phasons results in different regions of
localized electromagnetic energy at varying length scales. Then, scale
invariance is maintained only when a current operator $J_{na} (a =
\pm$) are the generators of momentum displacement an $x (q) (n
\epsilon k_+)$ within a
self similar space where $q^a_\mu \propto A^a_\mu$ ($a=\pm$ for
transverse, $a=0$ for long mode) is the momentum transfer of acoustic
modes correlated at $k_\mu$ such that $q_\mu (k_\mu) = k_\mu$ where,
like in a spin glass, $k_\mu$ is a localization scale. 
We
require $q_0 < q_1 < \cdots < Q_c$, which is the MD order parameter
with an algebra
\begin{equation}
[J_{n+}, J_{n^-}] = (k+n) J_0 (J_0 \propto q_0) 
\end{equation}
where $k$ is central charge with an effective charge $e^* = 1/(k+n)$.
$J_0$ can be seen as Magnetic Strips connected complementary QP
states.

\noindent\underbar{Order Parameter Symmetry}: One of the features of 
QP is the creation of a soliton excitation with an operator $Q
\epsilon$ SU(2) creating a scale invariant triad. This allows for
coherent transfer of charge triplet states similar to $He^3A$ being the zero mode
of conformal scaling and is seen as a gauge field causing charge solitons
to tunnel from one Peierls site to another. 
If the soft mode is
Raman active, then long range coherence requires a voltage modulation
of the pseudogap order parameter. Thus, if $C$ is the complex vector
space of the Higgs fields, and $A = (Q, \tilde Q) \epsilon$ SU(4) when
$Q, \tilde Q$ are
the SRO and LRO states while $G$ = SU(2) is the invariant subgroup,
for example, the electric or magnetic sectors of state space 
then ${\cal C} = (A/G) = SU(3)
\otimes U(1)_n$ where $n$ is the effective number of electrons
involved in the transition. A magnetic field $H$ will excite a phason
when the e.m. energy will propagate with quantum
interference at equivalent site caused by coherent dyon
oscillations. Topological quantization requires us to correlate
$(2+1)$ dim., space with a 2-dim dyon state and a 1-dim. gauge
invariant trajectory with phasons which restore scale invariance. Then
the second homotopy group measures a topological quantization of a
propagating excitation like a dyon moving coherently along different
layers keeping the transition amplitude from one region of $q(k)$ to
another SU(2) invariant. Thus
\begin{eqnarray}
\Pi{}_{_2} (C/SU(2)) &=& \Pi{}_{_2} (SU(2)) + \Pi{}_{_1} (U(1)_m)
\nonumber \\
&=& {\bf Z} \oplus  {\bf Z}_m 
\end{eqnarray}
Allowed states include $m=2$ and $m=4$ while dyons formed from EM
duality allows us to equivalence $m=1$ and $m=4$ states. Topological
quantization of the Quantum Peierls order parameter $Q_p$ requires
d-wave order coexisting with a Hall coefficient $\propto Q_v \propto
e^\ast = 1/(1+mn)$. A description of the pseudogap state with
co-existing order parameter is
\begin{equation}
Q \propto d \oplus d \oplus d \oplus e^\ast s
\end{equation}
when the fourth Casimir is a quantized charge showing a Fermi surface
anomaly. A phase coherent mode implies 
$\rho_n \propto \langle q^2_0 \rangle = k_{th}
(T) \propto e^\ast (T/T_g)$ where $T_g$ = pseudogap transition temperature
with a jump in $\rho_n \propto e^\ast$.  
Such a dependence should be seen in the pseudogap state and continued at
strong magnetic fields in the superconducting state. 

\noindent\underbar{Electronic Raman Mode (ERM) [2]} is, in our view, a
direct confirmation of coexisting order in the pseudogap
state. Experiments [2, Fig.1] can be thought of as measuring an
optical mode formed with bound charge singlet, spin triplet pairs in a
QP state with a spontaneously broken orbital symmetry with $L=1$ in
cuprates. A pair tunnels to a state of complementary chirality with a
dipolar interaction within a 2-phason model. Topological quantization
of phasons which are representations of the isotropy subgroup $H =$ SU
(2) requires a spin quantum number like a magnon allowing for coherent
charge
transfer. This is an example of Paramagnetic Meissner Effect where
magnetic Stripes like charge Stripes are expected. When a local pair
moves with the excitation of a d-wave magnetic diffuson [13], Raman
mode should be seen along [110] while when a surface mode (s-wave) is
excited the mode should be along [100] which is the d-wave gap at the
Fermi surface. The Raman continuum intensity is a spin susceptibility
determined by the thermal decoherence rate $1/\tau
\propto T$.  
%Clearly, NMR shift can be modulated by a factor $1/m$ due to
%scale invariance when the $m$ Raman mode is excited with $m$
%photons. Similarly, a uniform magnetio field causes (a) phase
%fluctuations from local rearrangements with $M(T) \propto HT^\propto
%(\propto \simeq 1)$ is a quantum critical exponent [~~~~] and (b)
%amplitude fluctuations arising from one ``length scale at an angular
%averaged $\langle \Delta (\theta) \rangle$ with overall scale
%invariance chensing $a (-dn/dE)$ factor [22]. 

\noindent\underbar{Magnetic relaxation} is a convincing demonstration of
a magnetic dyon in the pseudogap state seen in spin lattice
relaxation. An incident spin wave propagates with a coherent
absorption and emission of a soft lattice mode. 
A finite phase acquired by coherent electron lattice
scattering results in a finite displacement $Q^{-1}_c$ over a
localization length (or momentum) converting a spin wave to a phason
of a QP state. Since the propagation is phase
coherent, a spin wave excite a spin diffuson which is a quantum
inteference state with coherent exchange of a phason. Then the spin
lattice relaxation rate $1/T_1$ is given to within temperature
independent form factors by
\begin{equation}
1/T_1 \propto Q_+ Re {1 \over -i\omega + D (Q^2_c + Q^2_{th})} Q_- 
\end{equation}
where $Q_a = Q_p \oplus a Q_{th} (a=\pm)$ is a pairing amplitude correlation
momentum required for correlated momentum transfer and gauge
invariance and $Q^2_{th} \propto T$ is the phason order mode. Then
\begin{equation}
1/T_1T \propto T^\ast/(T + T^\ast) 
\end{equation}
with $T^\ast \sim T_c \propto \langle Q^2_c \rangle$ in magnetic
relaxation. In the superconducting state
$Q_c$ becomes the superconducting gap while the thermal localization
gap becomes irrelevant with $Q_{th} \to Q_h$ (coming from dipolar
energy) with a crossover at $Q_h
= Q_{th}$ or $H_{c2} (t) \propto t (t = 1 - T/T_c)$ for spin
fluctuation superconductors [18].  Similarly, magnetization fluctuations
are relevant when $Q_p \to Q_{th}$. In order for a phase coherent mode
in a superconductor we require a gauge transformation factor $\exp
(i\Phi/\Phi_0)$ being the same as $\exp (im\varphi)$ ($m=4$ for d-wave
order) which is the order parameter symmetry. In this way, a global
phase takes into account a conformal anomaly with $e^\ast =
(m/2\pi)$. A pseudogap transition is predicted at $T_g \sim
(1/e^\ast)T^\ast \sim$ 160K for $T^\ast \sim$ 103K while a low
temperature state of YBCO is predicted at temperatures $T < T_k \sim
(2/\pi) T^\ast \sim$ 65K [10].

%Actually, the diffusonic result for spin lattice relaxation is another
%version of Ampere's Law for a magnetic dipole current obtained by a
%motion of local electric flux lines within scale invariance
%states. These states transport a magnetic charge along the axis
%causing diamagnetism. Then $\chi (T) \propto \int dQ^2_\alpha /
%Q^2_{th} \propto |\ell n (T/T)|$ or $|\ell n (T/E)|$ for XY like
%transitions. A distinct calculation of specific heat and magnetic
%susceptibility [8] uses a Ginzburg Landau approach valid in the
%critical regime finding good agreement with experiment with pairing
%fluctuations for $T > T_c$.
%While the actual result for $M(t)$ which is the height of a spin echo
%phase [10] is known, our calculation shows the possibility of finding
%resonance at $1/T_1 \propto Q_c$ and at $m/2 T_1$ for $m=4$ for d-wave
%superconductors.  Moreover finding regious of superconductors with
%separate values of $m$ and allowing an electron to make a transition
%amongst any two implies for electron mediated transitions. 

Relaxation
rates at $1/T_1$ and $n/T_1$ when $n = mC_2$ for $m=2,3,4$ confirm
our belief in strong electron correlations among regions of different values
of phason momentum arising from disorder.
Magnetic stripes is a prediction of such a spin gap which is a
diamagnetic order parameter $Q_h$ implying a phase
coherent mode at magnetic fields when $L_h = Q^{-1}_h = cT^{(m)}_1$
($c=$ spin wave velocity) with $T_1^{(m)} = mT_1$ or $nT_1$
depending on experiments enabling local charges to restore symmetry.  

\noindent\underbar{Magnetic Diffuson (MD) Order}: An important
implication of a linear bias in the tunneling conductance is the
identification of a coherent dipolar coupling at $E_d \sim$ 42 meV
arising from a spin anomaly which we call a Paramagnetic Meissner
Effect. Then the operator $(Q \oplus iQ_{th})$ excites a mode
propagating forward in time with a phase coherent current $Q_v \hat z$
with the excitation of a harmonic oscillator mode at $Q_c$ which is
seen in NMR with an excess conductance
\begin{equation}
G(V) \propto Q_+ {1 \over -i\omega + DQ^2_v} \cdot Q_-  \propto (Q^2 -
Q^2_{th} (T)) \delta (V) 
\end{equation}
requiring, as usual, $Q^2_c (T) \propto T$ and $Q^2_v \propto V$ with
$G(V)$ measuring the superfluid component implying a s-wave order
parameter $Q_v$ with an energy source. Clearly, in our model, the
observation of an sharp peak in $G(V)$ at $V=0$ helps in
distinguishing charge dipolar interactions when the peak should not move
with the magnetic field for $k_h < k_{th}$ implying a
localization gap at the Fermi surface [14]. 
%Similarly, in the bulk mode, a bias increases the
%number of tunneling electrons with a coherent mode over the Coulomb
%gap implying [22,23] 
%\begin{equation}
%G(V) \propto (Q^2_c - Q^2_v)/(-i\omega_v + D (D^2 + Q^2_{th})),
%\omega_v = eV .
%\end{equation}
%This arises from a localized source of EH energy with a coherent made
%at dipolar energy restoring scaling. 

\noindent\underbar{Quantum Critical Phenomena} is seen as a magnetic
diffuson (MD) induced low energy modes with conformal scaling.  A
possibility of finding invariants $E_k$(read energy dispersion of low
energy states) consistent with our scaling hypothesis is obtained by
recalling the algebra for conserved currents when $J_{a} = k_x \oplus
ak_y$ which are the generators of SU(2) or SL(2,{\bf R}) requiring
\begin{equation}
E_k^2 \propto (k^2_x + k^2_y), |k^2_x - k^2_y|, (k_x - Q_p)^2 \cdot (k_y -
Q_p)^2, |k_x, k_y| > Q_p 
\end{equation}
which are the Onufrieva dispersions [20]. A disc like Fermi surface is
known. Then the Fermi energy $E_F$ is where the electronic states
change from localized to propagating. At $E_F$, an incident beam will
excite an electron across a localization length $L_c$ resulting in a
rearrangement of states with correlated absorption and emission of a
phasons.  At $E^+_F$, the states become propagating with $n > n_c$ and
$E_\alpha = n_c E$ (dipolar) and since the states are equally spaced
in momentum. Order parameter correlation in energy is obtained from
phasons with
\begin{equation}
\zeta (E) \propto {1 \over |E^2 - E^2_g|^(\nu/2)} 
\end{equation}
with $\nu=1$ for pseudogap state, $E_g = 0$ for $|E| < E_g$ and $E=0$
for Fermi 
liquid states with random potentials within Iterated Mean Field
(IMF). We expect a linear resistivity with a Drude $\rho (T) \propto
1/\tau (T) \propto T$ from phase decoherence. Evidently, the
verification of Eq.(10) with DOS $N(E) \propto \zeta^{-1} (E) \propto
E$ near the Fermi surface will verify a scale invariant order
parameter symmetry for d-wave order.

%of the form in Eq.(7) for the
%symmetries of QPS for short range, pseudogap or superconducting order,
%for the classical localization order and for an edge state order which
%is $m = 2$ for s-wave and $m=4$ for d-wave. Thus the diffusonic model
%makes specific verifiable predictions about order parameter symmetry
%from spin dynamics and the requirement of a coexisting QP and
%pseudogap order as a result of a self similar distribution of twinning
%sites.

An immediate prediction of diffusonic mode for a Drude relaxation rate
for infrared conductivity with scale symmetry restored y pseudogap at
$Q_c$ is [6]
\begin{equation}
\tau(\omega) \propto Re {1\over (-i\omega + Q_{th})} = \tau/(1+\omega^2
\tau^2), 1/\tau \propto Q^2_{th} \propto T  
\end{equation}
which is the relaxation rate of electrons to a global equilibrium
brought about when electrons at local equilibrium 
redistribute energies using a dipolar or phase decoherence term to
ensure that local charge anomalies are removed.

\noindent\underbar{Fermi surface anomaly}: 
MD is a dyon as a topologically quantized excitation arising from a
phason mediated pairing of chirally conjugate states at $\pm
\pi/m$. Since experiments measure phenomena at a set of length scales
FS will be determined by long range correlations max $(Q_v,Q_{th})$
with relevant symmetry. This is a likely explanation for an arc like
FS with a gap along symmetry directions arising from a coexisting
dipolar order [14,17]. This implies HTSC transition is a Quantum Critical
Phenomenon with a transition driven by low energy phason modes
inducing pairing correlations at an energy $T^\ast$ set by dyons with
$Q_v$ predicted to be correlated with symmetry gaps at the Fermi
surface.

\noindent\underbar{Stripes State}: Within our MD model with
coexisting order parameters, we are able to give an alternative
explanation for Stripes state. In the pseudogap state, exchange of
soft mode results in a
fluctuating charge gap due to phase fluctuations resulting in a
modulation of a QP order parameter $Q_{p\alpha} = (1/2) (1+ \alpha
k_x) (\alpha  = \pm)$
with our hypothesis $k_x = cx$. In order to explain domain sizes of
the magnitude seen in Carlson et al. [11] and Kivelson et al. [12], we set
$k_x \simeq 0.35$ since STM measurements which couple with states of
both chirality show ordering at $Q_+ \sim 0.7$ and $Q_- \sim 0.3$. We
require scale invariant states implying for modular $S$ duality a
requirement $Q_{pd} \to Q_{pd} + 1$ or average domain sizes
$\sim (2 \times 3 \times 4) a \sim 24a$ (linear dimension) and
superlattice sizes of length $(4 \times 5) a \sim 20a$ a = unit cell
spacing. In strong magnetic fields, superlattices of size $\sim (3
\times 4) a \sim 12a$ are predicted. Since we have charge symmetry,
within our hypothesis of a coexisting diamagnetic correlation length,
we expect order parameter modulations $Q_x \propto k_x (1 - dk_x)$ ($d
> 1$) implying minimal voltage fluctuations at $x \sim 1/8$, if one
has $c \sim 1.5$. $Q_x$ tracks the superconducting $T_c$ quite
accurately with our estimates in reasonable agreement with
phase diagram.

A longitudinal neutron beam excites a wave that coherently exchanges
an acoustic phonon causing neutron intensity oscillations at $\cos
(4Q_{h,z})$ with experiments showing $Q_{h} \sim$ 42 meV a dipolar
energy [16].  This is a direct verification of localization order and
is of importance in clarifying the charge spin dynamics of cuprates.
%Similarly, a monochromatic optical beam with a voltage $V$
%adjusted to ensure $k_v = k_h$ with both at orthogonality on the a-b
%plane will show Raman conductance oscillations with an intensity
%$\propto \cos (4E_n t) (E_v \propto Q_v)$ which is a verification a
%nested Fermi surface 
%with a minimum gap at $Q_c$ connecting states with $(\varphi + 2 pi/n),
%(\varphi - 2\pi/n)$ ($n=2$ for s-wave, $n=3$ for p-wave and $n=4$
%for d-wave order) and verified by a coherent phonon mode at $nE_c$
%with $E_c$ (d-wave $\sim$ 21 meV). 

%============= Figures and Captions ===============

%\newpage
\begin{figure}
\vskip 6cm 
\begin{center}
\includegraphics[scale=0.7]{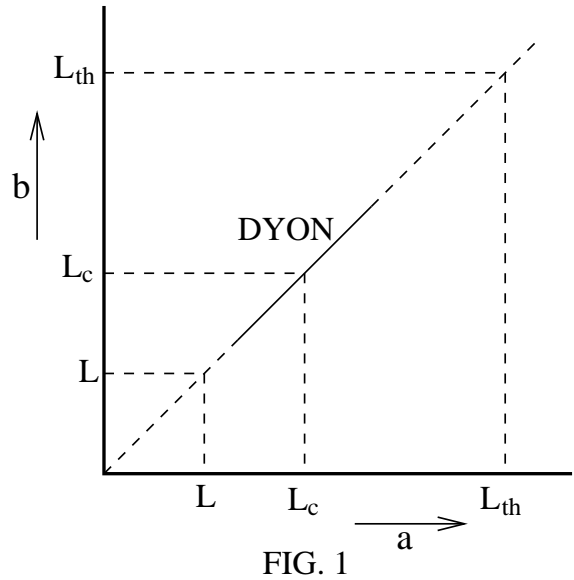}
\caption{Rubik's Cube representation of different length scales
corresponding to coexisting order parameters in the SU(2)
subspace. $L$ is a Peierls unit cell length, $L_{th}$ is a phase
decoherence length, $L_c$ is a scale invariant dyon. If $\ell$ is incident
wave  length,then an excitation at $L^2_c/\ell$ is generated with
energy invariants which are either circular (conventional) or
hyperbolic (non-conventional). Magnetic fluctuations are
accompanied by transitions of a dyon or dipole moment $e^\ast L_c$
shown as a solid line.} 
\end{center}
\end{figure}

\newpage

%============  References ===================

%%%%%\section*{\bf References}


\begin{thebibliography}{99}

\bibitem{1.}
J.G. Bednorz and K.A. M\"uller, Z. Phys. {\bf B64}, 189 (1986).
\bibitem{2.} 
C.M. Varma, Phys. Rev. Lett. {\bf 83}, 3538 (1999); Phys. Rev. {\bf
B55}, 14554 (1997); G. Blumberg et al., J. Phys. Chem. Solids {\bf
59}, 1932 (1988) and Phys. Rev. {\bf B53}, R11930 (1996). 
\bibitem{3.} 
D.J. Gross and M. M\'ezard, Nucl. Phys. {\bf B240} [FS12], 431 (1984);
N. Seiberg and E. Witten, Nucl. Phys. {\bf B426}, 19 (1994).
\bibitem{4.} 
S. Chakravarty, R.B. Laughlin, D.K. Morr and C. Nayak, Phys. Rev. {\bf
B63}, 094503I (2001); L. Balents, M.P.A. Fisher and C. Nayak,
Intl. Journ. of Mod. Phys. {\bf B12}, 1033 (1998). 
\bibitem{5.} 
A. Bianconi, et al., Phys. Rev. Lett. {\bf 76}, 3412 (1996). 
\bibitem{6.} 
T. Timusk and B. Statt, Rep. Progr. Phys. {\bf 62}, 61 (1999). 
\bibitem{7.} 
B. Banqu\'e, Y. Sidis, V. Hinkov, S. Pailh\'es, C.T. Lin, X. Chand and
Ph. Bourges, Phys. Rev. Lett. {\bf 96}, 19700 (2006). 
\bibitem{8.} 
C. Meingast, V. Pasler and P. Nagel, Phys. Rev. Lett. {\bf 86}, 1606
(2001); Phys. Rev. Lett. {\bf 89}, 229704 (2002). 
\bibitem{9.} 
W. J. Zheng, Phys. Rev. Lett. {\bf 83}, 3534 (1999). 
\bibitem{10.} 
V.F. Mitrovic, H.N. Bachman, W.P. Halperin, A.P. Reyes, P. Kahns and
W.G. Moulton, Phys. Rev. {\bf B66}, 014511 (2002). 
\bibitem{11.} 
E.W. Carlson, V.J. Emery, S.A. Kivelson, D. Orgad in the Physics of
Conventional and Unconventional Superconductors, ed. K.H. Bennemann
and J.B. Ketterson (Springes-Verlar); Q. Chen and J.R. Schrieffer,
Phys. Rev. {\bf B66}, 014512I (2002); Q. Chen et al., Phys. Rev. {\bf
B59}, 7083 (1997). 
\bibitem{12.} 
S.A. Kivelson, E. Fradkin, V. Oganesyan, I.P. Bindloss,
J.M. Tranqunada, A. Kapitalnik, C. Howald, Rev. Mod. Phys. {\bf 75},
1201 (2003). 
\bibitem{13.} 
D.R. Wake, Phys. Rev. {\bf B49}, 3641 (1994I). 
\bibitem{14.} 
A. Kaniegel et al., Nature Physics {\bf 2}, 447 (2006); M. Sutherland
et al., Phys. Rev. Lett. {\bf 94}, 147004 (2005). 
\bibitem{15.} 
J.C. Campuzano, M.R. Norman and M. Randeria in Physics of Conventional
and Unconventional Superconductors, ed. K.H. Bennemann and
J.B. Ketterson (Springer-Verlag). 
\bibitem{16.} 
G.M. Zhao, V. Kirtikar and D.E. Morris, Phys. Rev. {\bf B64}, 220506R
(2002); G.M. Zhao, cond-mat/0302566. 
\bibitem{17.} 
N.P. Armitage et al., Phys. Rev. Lett. {\bf 87}, 147003I (2001). 
\bibitem{18.} 
P. Bhattacharyya, Mod. Phys. Lett. {\bf B4}, 1177 (1990).
\bibitem{19.} 
P.W. Anderson, Science {\bf 235}, 1169 (1987); {\it ibid} {\bf 256},
1526 (1992); B. Edegger, V.N. Muthukumar, C. Gros and P.W. Anderson,
Phys. Rev. Lett. {\bf 96}, 207002 (2002). 
\bibitem{20.} 
F. Onufrieva, Physica {\bf 251}, 348 (1995).
\bibitem{21.}
F. Onufrieva and P. Pfeuty, Phys. Rev. {\bf B67}, 134525 (2003);
cond-mat/9807268. 
\bibitem{22.}
Ch. Renner et al., Phys. Rev. Lett. {\bf 80}, 149 (1998). 
\bibitem{23.}
N. Bergeal et al., cond-mat/0601265.
\bibitem{24.}
V. Ambegaokar,V. Eckern and G. Sch\"on, Phys. Rev. {\bf B30}, 6419
(1985).
%; D.J. Scalapino, Phys. Rev. Lett. {\bf 24}, 1052 (1970).  
\end{thebibliography}
\end{document}